\documentclass[fleqn,twoside]{article}
\usepackage{espcrc2}
\usepackage{amssymb}
\usepackage{amsmath}

\usepackage{graphicx}
\usepackage[figuresright]{rotating}

\newcommand{\stheta}{\ensuremath{\sin^22\theta_{13}}}
\newcommand{\BB}{$\beta$B}

\newcommand{\delCP}{\ensuremath{\delta_{\rm CP}}}

\newcommand{\He}{\ensuremath{^6{\mathrm{He}}}}
\newcommand{\Ne}{\ensuremath{^{18}{\mathrm{Ne}}}}

\title{\vspace*{-5mm}\flushright{\small CERN-PH-TH/2006-241}\\
\flushleft
Comparison of the CERN--MEMPHYS and T2HK neutrino oscillation
       experiments\thanks{Talk given at NOW 2006, 9--16 Sep 2006,
       Conca Specchiulla, Otranto, Italy}}

\author{Thomas Schwetz\address{CERN, Physics Department,%
        Theory Division, CH-1211 Geneva 23, Switzerland}}
       
\begin{document}

\begin{abstract}
\vspace*{-0.5pc}
In this talk I compare the physics potential of possible future
neutrino oscillation experiments from CERN to a Mt scale water
\v{C}erenkov detector at Fr\'ejus (MEMPHYS) and of the T2HK proposal in
Japan, where for the CERN experiments an SPL Superbeam and a
$\gamma=100$ Beta Beam are considered.
\vspace*{-0.5pc}
\end{abstract}

\maketitle

\section{INTRODUCTION}

Neutrino oscillation physics is now entering the era of precision
experiments~\cite{LBL}. The main aim of the upcoming generation of
experiments will be to establish a non-zero value of the mixing angle
$\theta_{13}$ or to push further the bound. On a time scale of 5 to 10
years decisions on a subsequent generation of high precision neutrino
oscillation facilities will have to been taken, and currently active
investigations on comparing various options are
performed~\cite{ISS}. Along these lines, in this talk I consider three
particular setups which are comparable in size and time scale, namely
two CERN based neutrino oscillation experiments consisting of a Beta
Beam (\BB)~\cite{BB} with $\gamma=100$ and a Superbeam
(SPL)~\cite{SPL}, as well as the phase II of the T2K experiment in
Japan (T2HK)~\cite{T2K}. All three configurations use a Mt scale water
\v{C}erenkov detector~\cite{Katsanevas}: MEMPHYS~\cite{memphys} at
Fr\'ejus or the Hyper-Kamiokande~\cite{Nakamura:2003hk} detector in
Japan. The main characteristics of the setups are displayed in
Tab.~\ref{tab:setups}. The results presented here are based on the
work~\cite{Campagne:2006yx}, where details on the calculations,
references, and more physics results and discussions can be found.

\begin{table}[tb]
  \begin{tabular}{lcc@{\quad}c}
  \hline
       & \BB & SPL & T2HK \\
  \hline
  Det.\ mass & 440~kt & 440~kt & 440~kt\\
  Baseline      & 130 km & 130 km & 295 km \\
  $\langle E_\nu \rangle$ [MeV] & 400 & 300 & 760 \\
  Time ($\nu/\bar\nu$) 
                & 5/5 yr & 2/8 yr & 2/8 yr \\
  Beam          & $5.8\,(2.2) \cdot 10^{18}$  & 4 MW & 4 MW\\
  Systematics   & 2--5\% & 2--5\% & 2--5\%\\
  \hline
  \end{tabular}
  \caption{Summary of default parameters used for the simulation of
  the \BB, SPL, and T2HK experiments. For the \BB\ the beam intensity
  is given in \He (\Ne) decays/yr.\label{tab:setups}}
\end{table}

\begin{figure*}[!tb]
  \centering 
  \includegraphics[height=7cm]{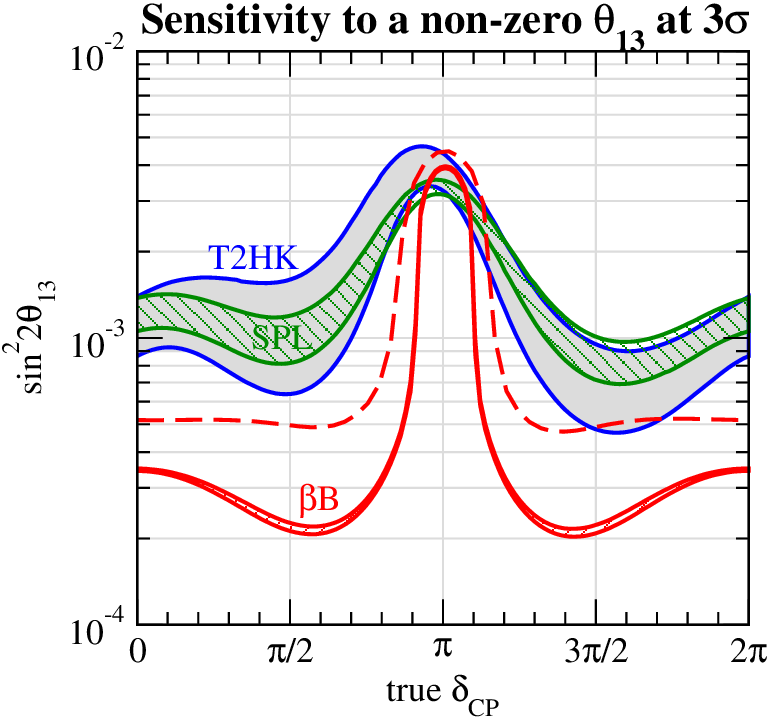}\qquad
  \includegraphics[height=7cm]{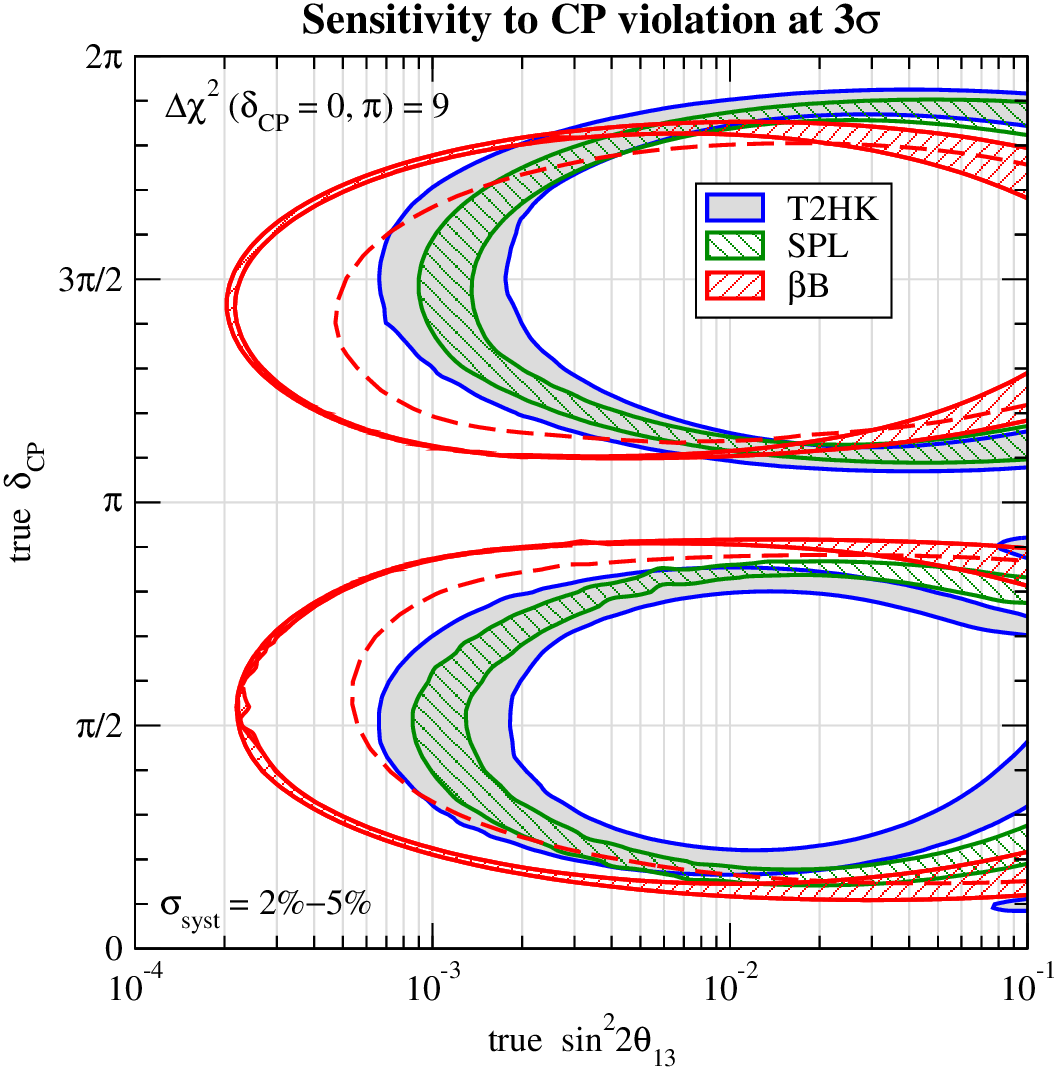}
  \vspace*{-.8cm}
  \caption{Left: $3\sigma$ sensitivity to $\stheta$ for \BB, SPL, and
  T2HK as a function of \delCP. Right: CPV discovery potential for
  \BB, SPL, and T2HK: for parameter values inside the ellipse-shaped
  curves CP conserving values of \delCP\ can be excluded at $3\sigma$
  $(\Delta\chi^2>9)$. The width of the bands corresponds to values for
  the systematical errors between 2\% and 5\%. The dashed curves show
  the sensitivity of the \BB\ when the number of ion decays/yr are
  reduced by a factor of two with respect to the values given in
  Tab.~\ref{tab:setups}.\label{fig:sens}}
\end{figure*}

\section{SENSITIVITY TO $\theta_{13}$ and CPV}

As performance indicators for the considered experiments we use the
potential to establish a non-zero value of $\theta_{13}$ and the
sensitivity to CP violation (CPV). The $\theta_{13}$ sensitivity is
shown in Fig.~\ref{fig:sens} (left) as a function of the value of the
CP phase \delCP: above the curves $\theta_{13} = 0$ can be excluded at
more than 3$\sigma$ CL (i.e., with $\Delta\chi^2 \ge 9$), whereas
Fig.~\ref{fig:sens} (right) shows the region in the $\theta_{13} -
\delCP$ plane where CPV can be established at 3$\sigma$.

One finds from these plots that SPL and T2HK perform rather
similar. In fact, the experimental setups of these two configurations
are similar (Superbeam technology, multi-MW beam power, detector,
$L/E_\nu$), the main differences being the shorter baseline of SPL
which implies lower neutrino energies (compare Tab.~\ref{tab:setups}),
as well as the use of an on-axis (off-axis) configuration for SPL
(T2HK). The lower energies for SPL imply that the cross section is
completely dominated by quasi-elastic (QE) scattering which allows a
good reconstruction of the neutrino energy, whereas at T2HK energies
non-QE events contribute significantly. On the other hand in the low
energy regime Fermi motion becomes important which limits energy
reconstruction. In our analysis we have taken these issues into
account by a migration matrix between true and reconstructed neutrino
energies based on detailed event simulations~\cite{Campagne:2006yx}.

Within our standard setup the \BB\ performs clearly better than the
Superbeams. For the \BB\ a crucial parameter is the total number of
ion decays. The conservative numbers from the EURISOL \BB\
studies~\cite{Lindroos}, which are two times smaler than our standard
values, lead to the sensitivites shown as dashed curves in
Fig.~\ref{fig:sens}.

\begin{figure}[!tb]
  \centering
   \includegraphics[width=0.48\textwidth]{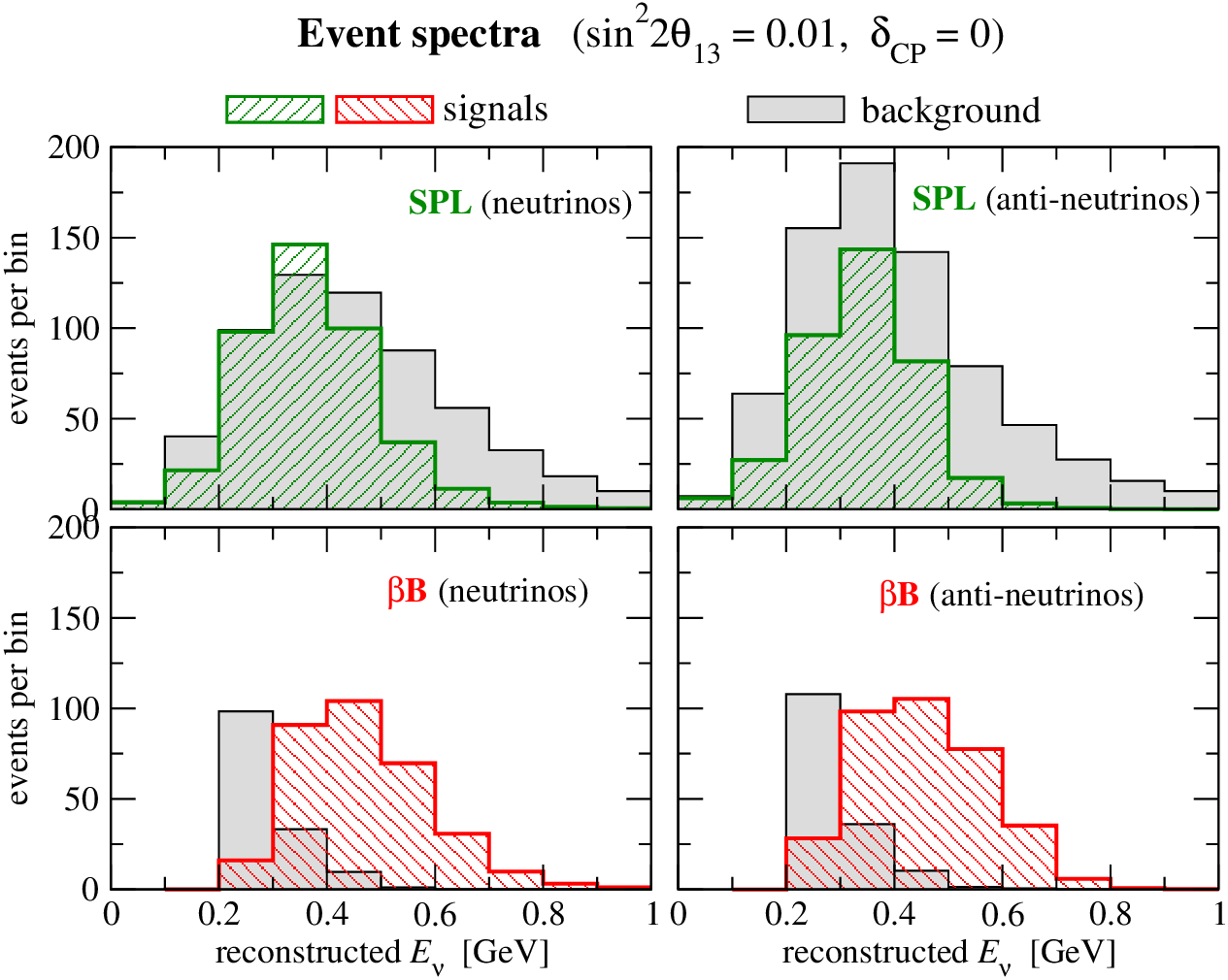}
   \vspace*{-1.2cm}
   \caption{Signal and background energy spectra for SPL (top) and
   \BB\ (bottom).\label{fig:spectra}}
\end{figure}

The widths of the curves in the figures shows the effect of varying
the (uncorrelated) systematical uncertainties on signal and
backgrounds between 2\% and 5\%. One can see that the \BB\ is
practically unaffected, whereas the Superbeam performances, and in
particular the one of T2HK, depend to some extent on the systematics.
The relevant systematic in this respect is the uncertainty on the
background. In the \BB\ the most important background comes from pions
produced mainly in NC $\nu_e/\bar\nu_e$ interactions which are
misidentified as muons. This background is efficiently reduced by
requiring to see the Michel electron from the muon decay. After
applying all the cuts roughly 300 background events remain. The reason
why an uncertainty on this number has so little impact for the \BB\ is
that the background has a very different shape than the signal. As
visible in Fig.~\ref{fig:spectra} it is peaked at low energies and
therefore spectral information makes the \BB\ very insensitive to the
systematical uncertainty. 

In contrast, for the Superbeams the background comes mainly from the
intrinsic $\nu_e/\bar\nu_e$ component of the beam and has a spectral shape
rather similar to the signal, as illustrated in Fig.~\ref{fig:spectra}
for the SPL. Therefore spectral information is not as efficient to
distinguish the background from the signal, as in the case of the \BB.

\section{COMMENTS ON DEGENERACIES}

In analyses of long-baseline experiments parameter degeneracies play
an important role. There are three types of
degeneracies~\cite{Barger:2001yr}, traditionally referred to as
intrinsic-, octant-, and sign($\Delta m^2_{31}$)-degeneracy. For the
experiments discussed here the intrinsic degeneracy is resolved by
spectral information but the degeneracies related to the octant of
$\theta_{23}$ and the neutrino mass hierarchy are present and lead to
ambiguities in the determination of $\theta_{13}$ and \delCP. However,
thanks to the fact that the matter effect is rather small (due to the
relatively short baselines) the degeneracies have very little impact
on the sensitivity to CPV. In other words, if the true parameter
values are CP violating also the parameter values of the degenerate
solutions will be CP violating. 

Moreover, as shown in Ref.~\cite{Huber:2005ep} atmospheric neutrino
data in Mt scale water \v{C}erenkov detectors as considered here can
be used to resolve degeneracies (see also
Ref.~\cite{Campagne:2006yx}). By combining long-baseline and
atmospheric data the mass hierarchy can be identified at $2\sigma$~CL
provided $\sin^22\theta_{13} \gtrsim 0.02-0.03$, although none of the
considered experiments has sensitivity to the hierarchy from
long-baseline data alone. Furthermore, atmospheric data provides
sensitivity to the octant of $\theta_{23}$, and if combined with the
$\nu_\mu$ disappearance chanel available in the Superbeam
experiments there is sensitivity to the octant for $|\sin^2\theta_{23}
- 0.5| \gtrsim 0.05$.

\section*{ACKNOWLEDGEMENTS}  

I thank J.-E.~Campagne, M.~Maltoni, and
M.~Mezzetto for collaboration on this topic. These results have been
obtained using the GLOBES~\cite{Globes} and NUANCE~\cite{Nuance}
software packages. The work of T.S.\ has been supported by the
$6^\mathrm{th}$~Framework Program of the European Community under a
Marie Curie Intra-European Fellowship at SISSA, Trieste, Italy.

\end{document}